\documentclass[letterpaper]{article} 
\usepackage{aaai2026}  
\usepackage{times}  
\usepackage{helvet}  
\usepackage{courier}  
\usepackage[hyphens]{url}  
\usepackage{graphicx} 
\usepackage{natbib}  
\usepackage{caption} 
\usepackage{algorithm}
\usepackage{algorithmic}

\usepackage{subfig} 
\usepackage{amsmath, amssymb, amsfonts} 
\usepackage{booktabs, makecell, multirow, tabularx} 
\usepackage{standalone}
\usepackage{xcolor}

\usepackage{microtype}
\usepackage{newfloat}
\usepackage{listings}
\DeclareCaptionStyle{ruled}{labelfont=normalfont,labelsep=colon,strut=off} 
\floatstyle{ruled}
\newfloat{listing}{tb}{lst}{}
\floatname{listing}{Listing}
\title{Multifaceted Scenario-Aware Hypergraph Learning for Next POI Recommendation}

\author {
    Yuxi Lin\textsuperscript{\rm 1},
    Yongkang Li\textsuperscript{\rm 2},
    Jie Xing\textsuperscript{\rm 1},
    Zipei Fan\textsuperscript{\rm 1,}\thanks{Corresponding author.}
}

\affiliations{
    \textsuperscript{\rm 1} School of Artificial Intelligence, Jilin University, China\\
    \textsuperscript{\rm 2} Informatics Institute, University of Amsterdam, The Netherlands\\
    
    \textrm{fanzipei@jlu.edu.cn}
}

\begin{document}

\maketitle

\begin{abstract}
Among the diverse services provided by Location-Based Social Networks (LBSNs), Next Point-of-Interest (POI) recommendation plays a crucial role in inferring user preferences from historical check-in trajectories. However, existing sequential and graph-based methods frequently neglect significant mobility variations across distinct contextual scenarios (e.g., tourists versus locals). This oversight results in suboptimal performance due to two fundamental limitations: the inability to capture scenario-specific features and the failure to resolve inherent inter-scenario conflicts. To overcome these limitations, we propose the \underline{\bf{M}}ultifaceted \underline{\bf{S}}cenario-\underline{\bf{A}}ware \underline{\bf{H}}yper\underline{\bf{g}}raph Learning method (MSAHG), a framework that adopts a scenario-splitting paradigm for next POI recommendation.
Our main contributions are:
(1) Construction of scenario-specific, multi-view disentangled sub-hypergraphs to capture distinct mobility patterns;
(2) A parameter-splitting mechanism to adaptively resolve conflicting optimization directions across scenarios while preserving generalization capability.
Extensive experiments on three real-world datasets demonstrate that MSAHG consistently outperforms five state-of-the-art methods across diverse scenarios, confirming its effectiveness in multi-scenario POI recommendation.\looseness=-1
\end{abstract}

\begin{links}
    \link{Code}{https://github.com/COCOMiss/MSAHG}
\end{links}

\section{Introduction}

Location-Based Social Networks~(LBSNs), such as Weibo, Foursquare, and Twitter, have become ubiquitous in modern digital life. By leveraging users’ check-in histories at Points-of-Interest (POI) and their social interactions, these platforms enable various personalized services, including friend~\cite{Li_hhgnn_cikm22,LiFYJDS23,li_tkdd_h3gnn} and next POI recommendation~\cite{dchl,Zhang_www_HKGNN}, enhancing user experience and driving platform monetization. Among them, next POI recommendation has attracted particular interest for its multifaceted value—ranging from improving user satisfaction to enabling targeted advertising and supporting urban mobility analysis~\cite{transferlearning,urban2vec}.\looseness=-1

\begin{figure}[t] 
    \centering
    \setlength{\fboxsep}{0pt}
    \setlength{\abovecaptionskip}{5pt}
    \setlength{\belowcaptionskip}{0pt}
    
    \subfloat[User type: local user and  tourist. \label{fig:a}]{%
        \includegraphics[width= \linewidth ]{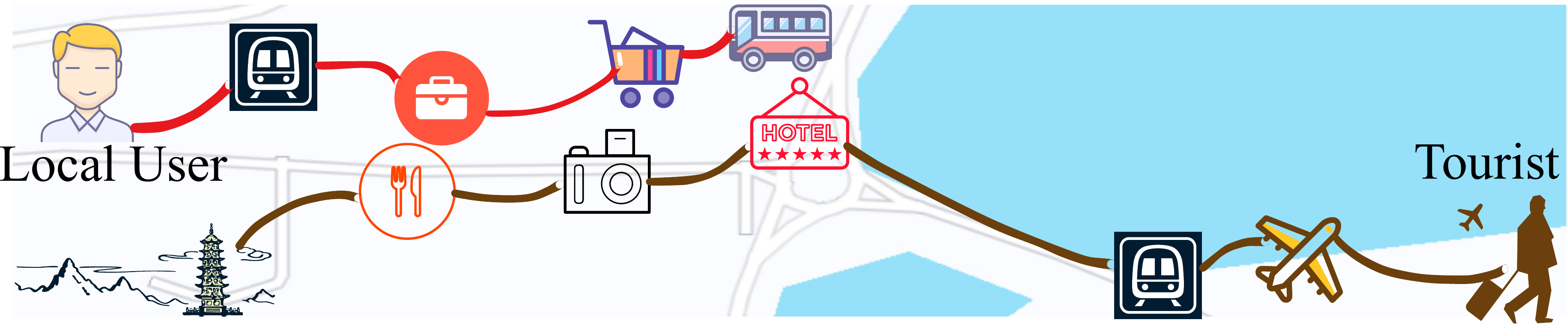}%
    }
    \vspace{0pt} 
    \subfloat[Temporal context: workday and weekend.\label{fig:b}]{%
        \includegraphics[width= \linewidth ]{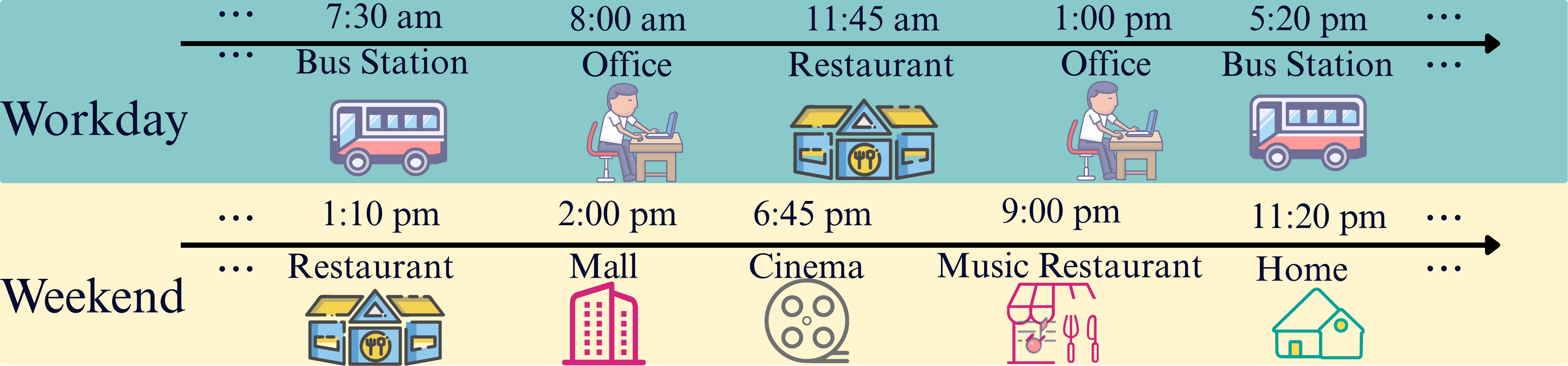}%
    }
    \vspace{0pt}
    \subfloat[Spatial region: suburban and downtown. \label{fig:c}]{%
        \includegraphics[width= \linewidth ]{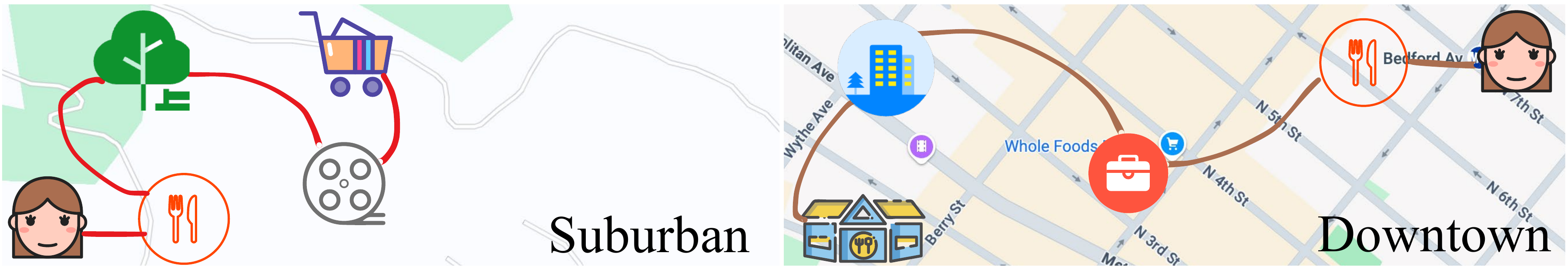}%
    }
\caption{Illustration of scenario-specific user preferences across three contextual dimensions in LBSNs, highlighting substantial behavioral differences across scenarios.
}
    \label{fig:introduction1}
\end{figure}

Researchers have proposed a wide range of approaches to tackle next POI recommendation, most notably sequential-based and graph-based methods. Among them, graph-based techniques have shown particular promise, leading to the emergence of hypergraph learning as a powerful framework in recent years, exemplified by models such as STHGCN~\cite{sthgcn} and DCHL~\cite{dchl}. Although these methods achieve strong performance in next POI recommendation, they often overlook a critical real-world requirement: recommendation systems should be capable of capturing distinct user preferences under varying contextual scenarios. 

To address this issue, we investigate the nature of contextual diversity in real-world recommendation environments. Based on empirical observations from Location-Based Social Networks (LBSNs), we categorize scenarios along three primary dimensions: user type (e.g., local vs. tourist), temporal context (e.g., weekday vs. weekend), and spatial region (e.g., downtown vs. suburban areas), as illustrated in~\figurename~\ref{fig:introduction1}.  For example, local users typically check in at locations relevant to their daily routines (e.g., offices, supermarkets), whereas tourists are more inclined toward leisure or landmark destinations. The same user may also exhibit different behaviors across time—frequenting transportation hubs and workplaces on weekdays while visiting restaurants or entertainment venues on weekends. In addition, spatial variations further influence mobility patterns: suburban users often remain within compact commercial areas, while downtown environments promote broader movement due to greater accessibility and POI diversity. These observations highlight the substantial behavioral differences exhibited by users under distinct contextual scenarios.

Motivated by these observations, we find that standard hypergraph structures and conventional methods struggle to capture the distinct trajectory patterns under diverse contextual scenarios. To address this limitation, we propose \underline{\bf{M}}ultifaceted \underline{\bf{S}}cenario-\underline{\bf{A}}ware \underline{\bf{H}}yper\underline{\bf{g}}raph Learning (MSAHG), which adopts a splitting-based approach to model each scenario individually: it constructs scenario-specific, multi-view disentangled sub-hypergraphs to capture distinct trajectory features. In parallel, it introduces a parameter splitting mechanism that dynamically resolves conflicting optimization directions across scenarios during training while maintaining generalization through shared parameters. This design overcomes a key limitation of conventional models—their inability to simultaneously accommodate multiple scenario-specific patterns without mutual interference. By explicitly modeling diverse contextual scenarios through sub-hypergraphs and adaptively splitting parameters with scenario conflicts, MSAHG effectively captures fine-grained mobility features across distinct real-world contexts.

Our main contributions are summarized as follows:

$\bullet$ We propose MSAHG, a splitting-based framework that explicitly models distinct scenarios—user types, time periods, and spatial regions—at both data and parameter levels.\looseness=-1

$\bullet$ We introduce two key techniques: (i) scenario-specific, multi-view disentangled sub-hypergraph construction to capture distinct mobility patterns, and (ii) a dynamic parameter-splitting mechanism to resolve conflicting optimization directions across scenarios while preserving generalization.

$\bullet$ We conduct extensive experiments on three real-world LBSN datasets. Results demonstrate that MSAHG consistently outperforms state-of-the-art baselines under diverse scenarios, validating its effectiveness in addressing the multi-scenario next POI recommendation task.

To the best of our knowledge, this is the first work that explicitly formulates and tackles the multi-scenario next POI recommendation problem, emphasizing the importance of scenario-aware modeling in real-world LBSNs.

\section{Related Work}
\subsection{Next POI Recommendation} 
Existing next POI recommendation approaches include sequential-based and graph-based methods. Sequential models~\cite{hst-lstm,Mobtcast} process trajectories as time series with spatio-temporal signals: DeepMove~\cite{deepmove} enhances RNN-based modeling with attention mechanisms, and MTNet~\cite{mtnet} segments trajectories to learn temporal-specific behavioral patterns. On the other hand, graph-based methods model user-POI interactions through spatial/semantic graphs\cite{SNPM,graph_transformer}. GETNext~\cite{getnext} constructs a global trajectory flow map to enrich POI embeddings within a Transformer\cite{transformer} framework. STGIN~\cite{stgin} captures context-specific preferences utilizing subgraphs, though it lacks explicit scenario definitions. More recently, hypergraph-based models have shown promise in capturing high-order relations in user trajectories\cite{LBSN2Vec,HGNN}. STHGCN~\cite{sthgcn} leverages hypergraph convolution to encode complex dependencies, while DCHL~\cite{dchl} disentangles heterogeneous trajectory features with multi-view hypergraphs. \emph{However, none of these methods explicitly account for the multifaceted scenario trajectory patterns.}

\subsection{Multi-scenario Recommendation}
While prior work has advanced next POI recommendation through sequential, graph, and hypergraph modeling, these methods largely operate under a single unified user behavior assumption. In contrast, multi-scenario recommendation aims to improve performance across diverse contexts by explicitly modeling scenario-level distinctions.
Recent approaches primarily focus on capturing inter-scenario similarities and differences within unified architectures. For example, STAR~\cite{star} adopts a star topology to decompose multi-scenario learning into a shared central network and multiple scenario-specific branches. M2M~\cite{m2m} introduces a meta-learning framework tailored for multi-advertising scenarios, incorporating meta units for cross-scenario correlation, attention mechanisms for scenario-specific dependency modeling, and task-specific towers for joint prediction. SAR-Net~\cite{sar-net} enhances user representation by integrating scenario/item features through dual attention modules, followed by scenario-specific transformations to extract context-aware signals.

\section{Preliminaries}
In this section, we first provide a detailed definition of the next POI recommendation task on LBSN, followed by an elaboration on our multifaceted scenario.

\subsection{Problem Formulation}
Let $U = \{u_1, u_2, \dots, u_M\}$ be the set of users,  $P = \{p_1, p_2, \dots, p_N\}$ the set of POIs, and $T = \{t_1, t_2, \dots, t_K\}$ the set of time slots, where $M$, $N$, and $K$ denote the numbers of users, POIs, and time slots, respectively. A POI check-in is represented as a triplet $\langle u_i, p_j, t_k \rangle$, where user $u_i$ visits POI $p_j$ at time $t_k$. Each POI $p$ is described by its geographical location $\langle \text{lat}, \text{lon} \rangle$. 
Given a user’s complete check-in history $Q_u = (q_{1}, q_{2}, \dots)$, where each trajectory sequence $q$ corresponds to a time interval (e.g., 24 hours), the goal is to predict the top-$k$ POIs that user $u$ is most likely to visit next, based on each individual trajectory sequence $q = (\langle u_i, p_1, t_1 \rangle, \langle u_i, p_2, t_2 \rangle, \dots, \langle u_i, p_n, t_n \rangle)$, from a set of $N$ candidate POIs.

\subsection{Multifaceted Scenario Definition}

\begin{figure}[t]
    \centering
    \includegraphics[width= \columnwidth]{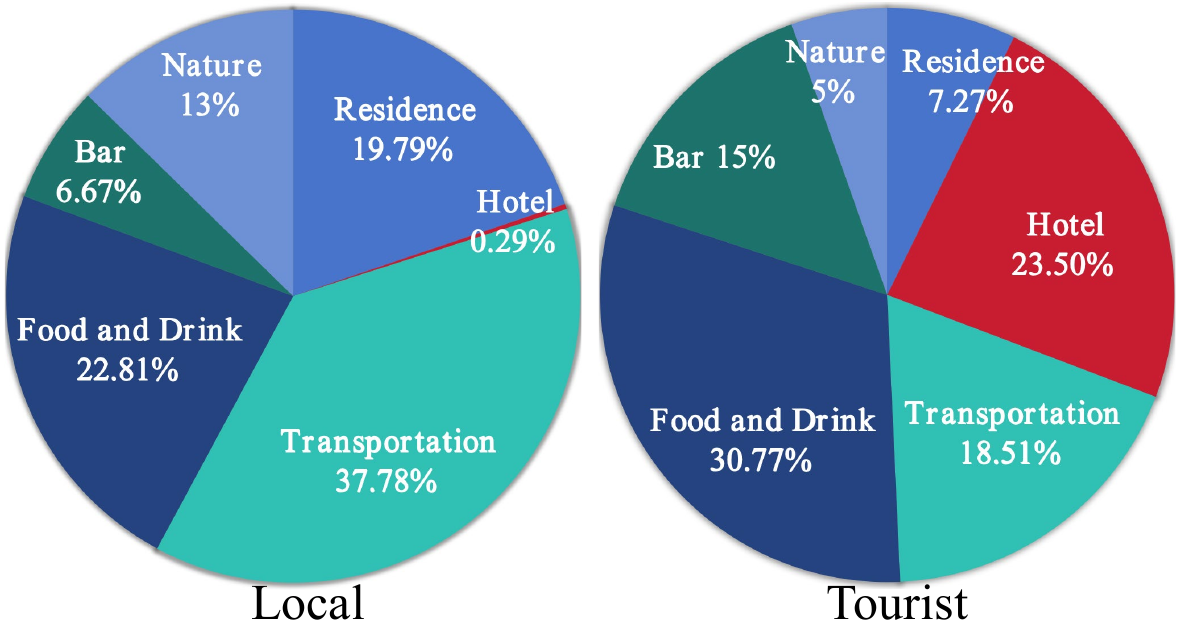}
    \caption{POI category distributions in NYC trajectories of local users and tourists. The marked differences, especially in Hotel, Residence, and Transportation categories, underline the necessity of scenario-aware modeling.}
    \label{fig:preliminaries_user_category}
\end{figure}

To capture diverse user behaviors under varying contexts—an example of which is illustrated in Figure~\ref{fig:preliminaries_user_category}—we extend the standard next POI recommendation task to a multifaceted scenario-aware setting. Specifically, each trajectory is categorized along three orthogonal scenario dimensions: user type, temporal context, and spatial region.\looseness=-1

$\bullet$  \textbf{User type}: Users are classified as \textit{local} or \textit{tourist} based on accommodation POI check-in frequency. A user is labeled as a \textit{tourist} if their accommodation check-ins (e.g., at hotels, resorts) exceed a threshold percentage of total check-ins (e.g., 5\%); otherwise, they are classified as \textit{local}.

$\bullet$ \textbf{Temporal context}: Trajectories are classified as \textit{weekday} or \textit{weekend} based on their final check-in timestamp.

$\bullet$ \textbf{Spatial region}:Trajectories are categorized as \textit{downtown} or \textit{suburban} based on their last location of check-in: within 10 km of city center for downtown, otherwise suburban.\looseness=-1

Each trajectory is uniquely assigned to one category per dimension, forming a combined scenario (e.g., ``local \& weekend \& suburban''). The task of multifaceted scenario-aware POI prediction is to recommend the next POI within these composite scenarios, aiming to improve performance across heterogeneous scenarios.\looseness=-1

\section{Methodology}

\begin{figure*}[ht]
    \centering
    \includegraphics[width=\textwidth]{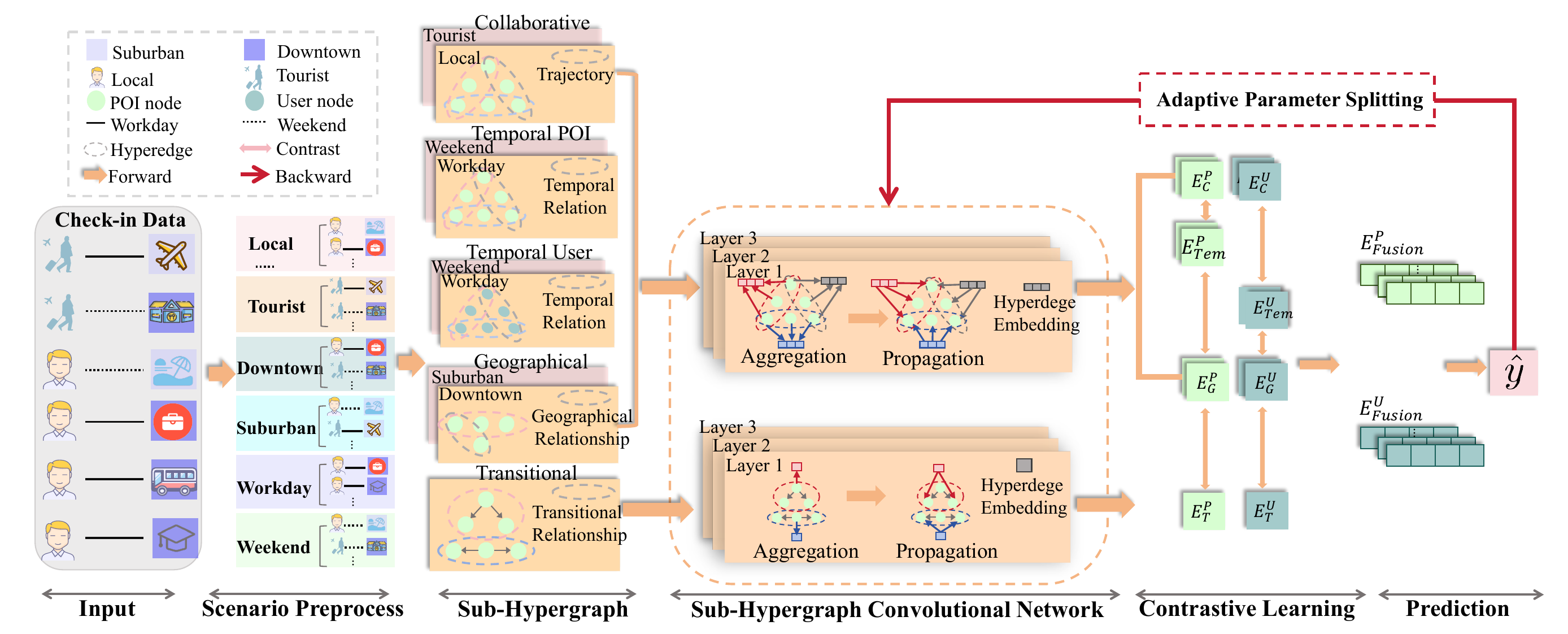}
    \caption{The overall framework of \underline{\bf{M}}ultifaceted \underline{\bf{S}}cenario-\underline{\bf{A}}ware \underline{\bf{H}}yper\underline{\bf{g}}raph Learning (MSAHG). User trajectories are categorized into three scenario dimensions: user type, temporal context, and spatial region. Based on these and relational patterns, eight sub-hypergraphs are constructed and processed via multi-layer hypergraph convolution. A parameter splitting mechanism handles conflicting scenario signals. Final embeddings are learned through contrastive learning and fused for next POI prediction.
    }
    \label{fig:model}
\end{figure*}

In this section, we introduce MSAHG, a framework for multi-scenario next POI recommendation that explicitly models trajectory features across scenarios (see Figure~\ref{fig:model}). Within each scenario, user behaviors tend to be consistent, while mixing data across scenarios may introduce noise and harm performance. Additionally, joint optimization over heterogeneous scenarios can lead to conflicting gradients for shared parameters. To address these challenges, MSAHG employs two components: a multi-view disentangled sub-hypergraph to capture intra-scenario mobility patterns, and an adaptive parameter splitting mechanism to resolve inter-scenario conflicts.\looseness=-1

\subsection{Sub-Hypergraph Construction}

To model heterogeneous trajectory patterns across scenarios, MSAHG constructs separate sub-hypergraphs for each scenario to better capture their scenario-specific semantics. As illustrated in Figure~\ref{fig:model} (Sub-Hypergraph Construction section), we generate eight types of sub-hypergraphs from four complementary views: collaborative, temporal user, temporal POI, and geographical.

\textbf{User type view.} We divide users into \textit{locals} and \textit{tourists} based on their check-in histories (Figure~\ref{fig:a}). Locals tend to follow routine daily mobility patterns, while tourists typically visit landmarks and restaurants with irregular time and location preferences. Due to these behavioral differences, we construct separate collaborative sub-hypergraphs for each user type, where nodes represent POIs and hyperedges represent individual user trajectories (Figure~\ref{fig:model}, first sub-hypergraph).

\textbf{Temporal view.} User behaviors also differ between weekdays and weekends (Figure~\ref{fig:b}). To capture temporal scenario patterns, we build two types of sub-hypergraphs: temporal-user and temporal-POI. We divide each day into 48 equal time intervals, and use these as hyperedges. In the temporal-user sub-hypergraph, nodes represent users; in the temporal-POI sub-hypergraph, nodes are POIs (Figure~\ref{fig:model}, second and third sub-hypergraph).

\textbf{Geographical view.} Trajectory patterns vary between urban and suburban areas due to differences in POI types and spatial distribution (Figure~\ref{fig:c}). We therefore construct separate geographical sub-hypergraphs based on region. In this view, hyperedges represent geographical neighborhoods~(POIs within a specific distance threshold), and nodes are POIs (Figure~\ref{fig:model}, fourth sub-hypergraph).

\textbf{Transitional view.} We model global POI transition patterns using a single transitional hypergraph shared across all scenarios. Nodes represent POIs, and hyperedges capture directed transitions between consecutive POIs in user trajectories. Since transition patterns are less scenario-dependent, we do not split this view into sub-hypergraphs (Figure~\ref{fig:model}, fifth sub-hypergraph).

\subsection{Sub-Hypergraph Convolutional Networks}
To obtain informative POI and user representations from the multi-view sub-hypergraphs, we apply a two-step sub-hypergraph convolution~(node-hyperedge-node), as shown in Figure~\ref{fig:model}. Inspired by the DCHL framework~\cite{dchl}, a residual connection is employed to mitigate over-smoothing. All sub-hypergraphs are undirected and scenario-specific, while the transitional hypergraph is directed and shared across scenarios.\looseness=-1

\subsubsection{Message Propagation in Sub-Hypergraphs}

Given an undirected sub-hypergraph \(\mathcal{G}_{\text{und}} = (\mathcal{V}, \mathcal{E}_{\text{und}})\) with node set \(\mathcal{V}\) and hyperedge set \(\mathcal{E}_{\text{und}}\), we initialize node embeddings stochastically. The node representations are updated through the following two operations, where \(\mathbf{m}_e\) denotes the hyperedge embedding:

$\bullet$  \textbf{Aggregation}: Node embeddings within a hyperedge \(e\) are aggregated to generate an intermediate message:
    \begin{equation}
    \mathbf{m}_e = \text{AGG}\left( \left\{ \mathbf{v}_i \mid v_i \in e \right\} \right)
        \label{eq:1}
    \end{equation}
    where \( \text{AGG}(\cdot) \) represents a node-to-hyperedge aggregation function (e.g., mean pooling).

$\bullet$  \textbf{Propagation}: Messages from the hyperedges that related to node \(\mathbf{v}\) are propagated to refine its representation:
    \begin{equation}
        \overline{\mathbf{v}}_i = \text{PROP}\left( \{ \mathbf{m}_e \mid e \in \mathcal{E}_{und} \} \right)
    \end{equation}
  
With residual connections across \( L \) layers, the final node embedding is computed as:
\begin{equation}
    \mathbf{v}_{final} = \frac{1}{L+1} \sum_{\ell=0}^L \left( \overline{\mathbf{v}}^{(\ell)} + \overline{\mathbf{v}}^{(\ell-1)} \right)
\end{equation}

For the directed transitional hypergraph \( \mathcal{G}_{\text{dir}} \), aggregation incorporates only source node embeddings into hyperedges, while propagation delivers hyperedge embeddings exclusively to target nodes.

\subsubsection{User \& POI Embedding} To obtain unified User and POI representations, we fuse scenario-specific embeddings that obtained from the sub-hypergraph convolutional networks. Consider a user visiting a downtown restaurant on Thursday: MSAHG integrates 1) collaborative embeddings from the local-user hypergraph, 2) temporal embeddings from the workday hypergraph, and 3) geographical embeddings from the downtown hypergraph (Figure~\ref{fig:model}). Note that transitional view embeddings are derived directly without sub-hypergraphs. \looseness=-1

User embeddings are derived from POI embeddings via the transposed incidence matrix \( H_C^T \) of the collaborative hypergraph. For collaborative (\(E_C^U\)), transitional (\(E_T^U\)), and geographical (\(E_G^U\)) views:
\begin{equation}
E_X^U = H_C^T \cdot E_X^P \quad (X \in \{C, T, G\})
\label{eq:user_embedding}
\end{equation}
where \(E_X^P\) denotes POI embeddings from view \(X\).
While, the user embedding~\( E_\text{Tem}^U \) can be directed obtained from temporal user view sub-hypergraphs, as their nodes are users.
To adaptively fuse these user embeddings, we introduce view-specific gating mechanisms:
\begin{equation}
E_F^U = \lambda_{\text{Tem}} E_\text{Tem}^U + \lambda_C  E_C^U + \lambda_T E_T^U + \lambda_G E_G^U
\end{equation}
where gating weights \(\lambda_X = \sigma(E_X^U \mathbf{W}_X)\) employ Sigmoid activation \(\sigma\) and view-specific trainable weights \(\mathbf{W}_X \in \mathbb{R}^d\). For POI representations,  we employ a simple summation of different view embeddings.

\subsection{Contrastive Learning}

To capture cooperative associations across multi-view embeddings, we employ contrastive learning with InfoNCE loss. For user representations, we define pairwise contrastive loss between views \(v_1\) and \(v_2\) as:
\begin{equation}
\mathcal{L}_{v_1,v_2}^U = -\frac{1}{|\mathcal{U}|} \sum_{u \in \mathcal{U}} \log \frac{\exp\left( s(\mathbf{e}_{v_1}^u, \mathbf{e}_{v_2}^u) / \tau \right)}{\sum_{u' \in \mathcal{U}} \exp\left( s(\mathbf{e}_{v_1}^u, \mathbf{e}_{v_2}^{u'}) / \tau \right)}
\label{eq:pairwise}
\end{equation}
where \( s(\cdot,\cdot) \) denotes cosine similarity, \(\tau\) is a temperature hyperparameter, and \(\mathbf{e}_v^u\) represents user's embeddings from view \(v\).

For each scenario, the scenario-specific loss is computed by summing its contrastive loss and POI recommendation loss. The user contrastive loss is obtained by summing losses across all pairs of views:

\begin{equation}
\begin{gathered}
\mathcal{L}_{\text{con}}^U = \sum_{\substack{v_1, v_2}} \mathcal{L}_{v_1,v_2}^U \\
\left( \text{where } v_1 < v_2 \text{ and } v_1, v_2 \in \{\text{Tem}, C, G, T\} \right)
\end{gathered}
\label{equation:contrastive2}
\end{equation}

Similarly, the contrastive loss for POI representations is defined as \( \mathcal{L}_{con}^P \). The final scenario-specific loss is:
\begin{equation}
\mathcal{L}_{final} = \lambda(\mathcal{L}_{con}^U + \mathcal{L}_{con}^P) + (1-\lambda)\mathcal{L}_{Rec}
\label{equation:contrastive3}
\end{equation}
where \(\lambda\) balances multi-view cooperation against recommendation accuracy, as illustrated in Figure~\ref{fig:model}.

\subsection{Adaptive Parameter Splitting}

\begin{figure}[t]
    \centering
    \includegraphics[width=1\columnwidth]{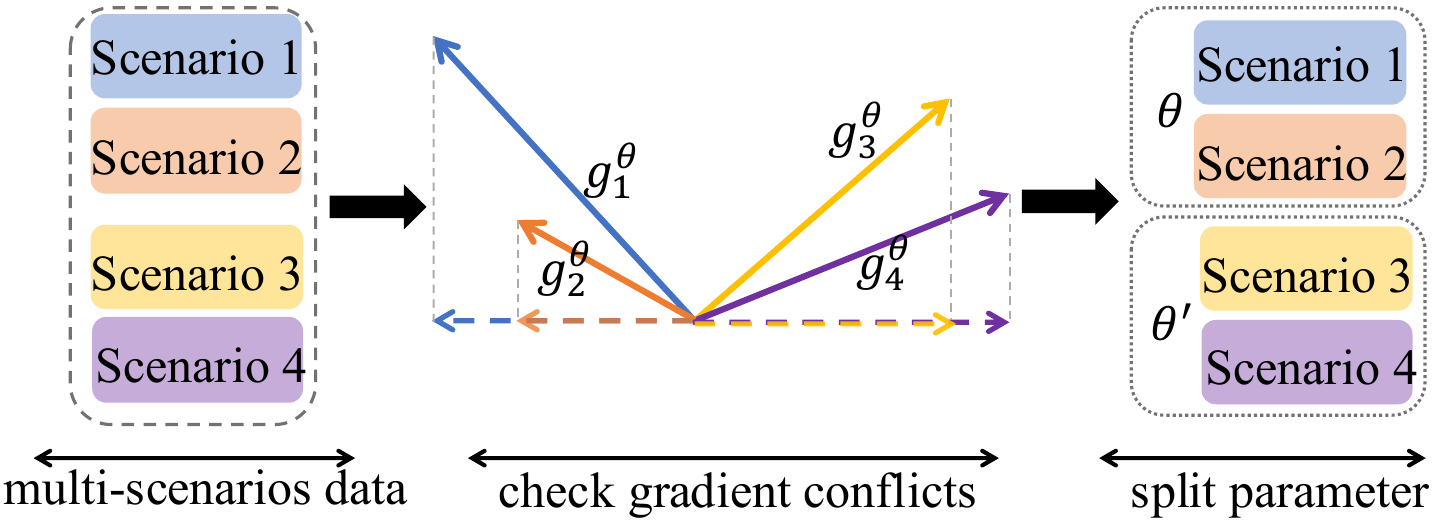}
    \caption{Overview of adaptive parameter splitting. For each parameter \( \theta \), gradients from different scenarios are compared. When conflicting directions are detected, \( \theta \)is duplicated, and scenarios are grouped to use either \( \theta \) or the new copy \( \theta' \).}
    \label{fig:split}
\end{figure}

In multi-scenario settings, shared parameters may receive conflicting gradient signals across scenarios, hindering generalizable representation learning. To address this, we propose an adaptive parameter splitting mechanism that selectively duplicates conflicting parameters during training.
Let \( \theta \) denote an individual parameter in the model, and \( N \) the number of scenarios. For each 
\( \theta \) , we compute the gradients of each scenario’s loss, normalize them, and calculate the cosine similarities among scenarios. For example, the similarity between scenario $i$ and scenario $j$ is calculated with equation~\ref{eq:splitting1}:\looseness=-1
\begin{equation}
s^\theta_{i,j} = {g}^\theta_i \cdot {g}^\theta_j
\label{eq:splitting1}
\end{equation}
where \( 0 \leq i, j < N \), and \( {g}^\theta_i \), \( {g}^\theta_j \) represent the normalized gradients of scenario \( i \) and scenario \( j \) with respect to the parameter \( \theta \), respectively. A negative similarity \( s^\theta_{i,j} <0 \) indicates that the two scenarios are conflicting on \(\theta\), suggesting that a single shared parameter cannot accommodate both. 
In such cases, we split the conflicting parameter \( \theta \) into two versions: the original \( \theta \) and a new copy \( \theta' \). 
Scenarios are clustered based on gradient similarity, and each group uses the corresponding parameter during forward and backward passes. This operation is applied independently per parameter and only to those that have not been split before. Once split, a parameter becomes scenario-specific and remains tied to its assigned group.
To reduce overhead, conflict checking and splitting are performed periodically (e.g., every fixed number of epochs). Although gradients are evaluated for all parameters, only a few typically require splitting. As a result, model size increases modestly and stays well below twice the original parameter count.
This mechanism supports fine-grained, scenario-aware learning while preserving parameter efficiency, as illustrated in Figure~\ref{fig:split}.

\section{Experiments}
In this section, we conduct comprehensive experiments to evaluate our proposed MSAHG model. Specifically, we first introduce the experimental setup, including the datasets, baselines, evaluation metrics, and implementation details. Then, we demonstrate the overall performance of our model in terms of multi-scenario recommendation and  ablation studies.\looseness=-1

\subsection{Experimental Setup}

\subsubsection{Datasets}
We evaluate MSAHG on three publicly accessible real-world datasets: Gowalla, NYC, and TKY. The NYC and TKY datasets are derived from Foursquare\cite{fousquare}; The Gowalla dataset\cite{gowalla} contains check-ins spanning from February 2009 to October 2010, covering California and Nevada. The details of datasets and categorizing data into multiple scenarios are available in Appendix.

\begin{table*}[!ht]
    \centering
    \renewcommand\arraystretch{0.95}
    \resizebox{\linewidth}{!}{
    \setlength{\tabcolsep}{0.6mm}{ %
    \begin{tabular}{@{}lcccccccccccccccc@{}}
        \toprule
        & & \multicolumn{5}{c}{Gowalla} & \multicolumn{5}{c}{NYC} & \multicolumn{5}{c}{TKY} \\
        \cmidrule(lr){3-7} \cmidrule(lr){8-12} \cmidrule(lr){13-17}
        \textbf{Model} & \textbf{Scenario} & \textbf{ACC@1} & \textbf{ACC@5} & \textbf{ACC@10} & \textbf{ACC@20} & \textbf{MRR} & \textbf{ACC@1} & \textbf{ACC@5} & \textbf{ACC@10} & \textbf{ACC@20} & \textbf{MRR} & \textbf{ACC@1} & \textbf{ACC@5} & \textbf{ACC@10} & \textbf{ACC@20} & \textbf{MRR} \\
        \midrule
        & Local & 0.1314 & 0.2758 & 0.3369 & 0.3934 & 0.2014 & \underline{0.2020} & \underline{0.3762} & 0.4237 & 0.4583 & \underline{0.2809} & \underline{0.2159} & 0.3878 & 0.4573 & 0.5203 & 0.2978 \\
        & Tourist & \textbf{0.1079}& 0.2349 & 0.2999 & 0.3623 & 0.1714 & 0.1298 & 0.2576 & 0.3056 & 0.3398 & 0.1886 & \textbf{0.1679}& 0.3137 & 0.3762 & 0.4367 & \textbf{0.2387}\\
        HSTLSTM & Downtown & 0.1405 & 0.2909 & 0.3558 & 0.4163 & 0.2122 & 0.2065 & 0.3731 & 0.4276 & 0.4650 & 0.2816 & \underline{0.2334}& 0.4236 & 0.5002 & 0.5626 & 0.3229\\
        & Suburban & 0.1102 & 0.2391 & 0.3019 & 0.3614 & 0.1746 & 0.1454 & 0.2877 & 0.3327 & 0.3658 & 0.2107 & 0.1785 & 0.3227 & 0.3828 & 0.4452 & 0.2489 \\
        & Workday & 0.1366 & 0.2846 & 0.3512 & 0.4133 & 0.2085 & 0.1723 & 0.3295 & 0.3793 & 0.4152 & 0.2439 & 0.2118 & 0.3813 & 0.4497 & 0.5121 & 0.2927 \\
        & Weekend & 0.0820 & 0.1901 & 0.2471 & 0.3023 & 0.1373 & 0.0240 & 0.0524 & 0.0689 & 0.0778 & 0.0394 & 0.0728 & 0.1715 & 0.2275 & 0.2897 & 0.1239 \\
        \midrule
        & Local & 0.1343 & 0.2984 & 0.3644 & 0.4242 & 0.2122 & 0.1603 & 0.3311 & 0.3871 & 0.4268 & 0.2382 & 0.1756 & 0.3791 & 0.4575 & 0.5232 & 0.2698 \\
        & Tourist & \underline{0.1074}& 0.2435 & 0.3066 & 0.3699 & \underline{0.1751}& 0.1178 & 0.2322 & 0.2827 & 0.3272 & 0.1722 & \underline{0.1455}& \underline{0.3155}& \underline{0.3861}& 0.4411 & 0.2254 \\
        DeepMove & Downtown & 0.1441 & 0.3114 & 0.3807 & 0.4402 & 0.2238 & 0.1544 & 0.3078 & 0.3606 & 0.3950 & 0.2235 & 0.1806 & 0.3897 & 0.4727 & 0.5392 & 0.2773 \\
        & Suburban & 0.1096 & 0.2505 & 0.3131 & 0.3758 & 0.1789 & 0.1305 & 0.2660 & 0.3192 & 0.3648 & 0.1944 & 0.1574 & 0.3406 & 0.4114 & 0.4714 & 0.2428 \\
        & Workday & 0.1397 & 0.3067 & 0.3760 & 0.4393 & 0.2199 & 0.1436 & 0.2900 & 0.3449 & 0.3882 & 0.2116 & 0.1730 & 0.3732 & 0.4509 & 0.5145 & 0.2658 \\
        & Weekend & 0.0772 & 0.1875 & 0.2420 & 0.3008 & 0.1332 & 0.0168 & 0.0440 & 0.0608 & 0.0797 & 0.0307 & 0.0751 & 0.1762 & 0.2282 & 0.2808 & 0.1269 \\
        \midrule
        & Local & \underline{0.1556}& \textbf{0.3385} & \textbf{0.4104} & \textbf{0.4768} & \textbf{0.2432} & 0.1782 & 0.2970 & 0.3762& \underline{0.5149}& 0.2398 & 0.1416 & 0.3625 & 0.4969& \textbf{0.6247} & 0.2479\\
        & Tourist & 0.0785 & 0.1983 & 0.2438 & 0.3058 & 0.1329 & 0.0714 & 0.2143 & 0.3571 & 0.4464 & 0.1519 & 0.0213 & 0.1277 & 0.2128 & 0.3404 & 0.0771 \\
        GETNext & Downtown & \textbf{0.1929} & \textbf{0.4180} & \textbf{0.5145} & \textbf{0.5627} & \textbf{0.2968} & 0.1000& 0.2000& 0.3000& 0.5000& 0.1600& 0.1511 & 0.4106& 0.5668& \textbf{0.7162} & 0.2753\\
        & Suburban & 0.1262 & 0.2816 & 0.3398 & 0.4107 & 0.2011 & 0.1460 & 0.2774 & \underline{0.3796}& \underline{0.4891}& 0.2155 & 0.1207 & 0.2783 & 0.3781 & 0.4741 & 0.1978 \\
        & Workday & 0.1494 & 0.3274 & 0.3980 & 0.4651 & 0.2337 & 0.1450 & 0.2748 & 0.3969 & 0.5115 & 0.2157 & 0.1438 & 0.3598 & 0.4912& \textbf{0.6259} & 0.2480\\
        & Weekend & \underline{0.1031}& \textbf{0.2422}& \textbf{0.2915}& \textbf{0.3498}& \textbf{0.1713}& 0.1154 & 0.2308 & 0.2308 & 0.3846 & 0.1719 & 0.1223 & \textbf{0.3476}& \textbf{0.4871} & \textbf{0.5923} & \underline{0.2304}\\
        \midrule
        & Local & 0.1433 & 0.2854 & 0.3505 & 0.3991 & 0.2115 & 0.1730 & 0.3460 & 0.4010 & 0.4382 & 0.2533& \textbf{0.2283}& \underline{0.4409}& \underline{0.5221}& 0.5902& \textbf{0.3271} \\
        & Tourist & 0.0543 & 0.1793 & 0.2554 & 0.3370 & 0.1168 & \underline{0.1841}& \underline{0.3899}& \underline{0.4296}& \underline{0.4657}& \underline{0.2663}& 0.1392 & 0.2840 & 0.3648& \underline{0.4422}& 0.2073\\
        STHGCN & Downtown & 0.1399 & 0.2762 & 0.3479 & 0.3864 & 0.2022 & \underline{0.2378}& 0.4483& 0.5263& 0.5692& 0.3360& \textbf{0.2459}& \textbf{0.4879} & \textbf{0.5774} & \underline{0.6484}& \textbf{0.3577} \\
        & Suburban & \underline{0.1368}& 0.2840 & 0.3483 & 0.3972 & \underline{0.2074}& \underline{0.1592}& \underline{0.3235}& 0.3667 & 0.4095 & \underline{0.2337}& \textbf{0.2148}& \textbf{0.4151} & \textbf{0.4983} & \textbf{0.5641} & \textbf{0.3082} \\
        & Workday & 0.1530 & 0.3141 & 0.3849 & 0.4348 & 0.2302 & \underline{0.1856} &0.3789 &0.4293 &0.4663 &\underline{0.2708} &  \textbf{0.2361} & \underline{0.4615} & \underline{0.5471} & 0.6124 & \textbf{0.3405} \\
        & Weekend & \textbf{0.1130} & 0.2218 & 0.2762 & 0.3222 & \underline{0.1647} & \underline{0.1423} & \underline{0.2397} & \underline{0.2996} & \underline{0.3390} & \underline{0.1971} & \textbf{0.1650} & \underline{0.3363} & \underline{0.4164} & \underline{0.4892} & \textbf{0.2478} \\
        \midrule
         & Local & 0.1364 & 0.3109 & 0.3818 & 0.4537 & 0.2203 & 0.1463 & 0.3555 & \underline{0.4244} & 0.4854 & 0.2408 & 0.1190 & 0.3058 & 0.3896 & 0.4627 & 0.2072 \\
         & Tourist & 0.0752 & \underline{0.2585} & \underline{0.3104} & \underline{0.3795} & 0.1569 & 0.1029 & 0.2111 & 0.2696 & 0.3031 & 0.1559 & 0.0658 & 0.2088 & 0.2735 & 0.3477 & 0.1391 \\
         DCHL & Downtown & 0.1477 & 0.3372 & 0.4070 & 0.4760 & 0.2361 & 0.1945 & \underline{0.4774} & \underline{0.5665} & \underline{0.6486} & \underline{0.3205} & 0.1405 & 0.3530 & 0.4410 & 0.5184 & 0.2398 \\
         & Suburban & 0.1247 & \underline{0.2929} & \underline{0.3597} &\textbf{0.4323} & 0.2056 & 0.1237 & 0.2937 & 0.3545 & 0.4041 & 0.2017 & 0.1010 & 0.2655 & 0.3448 & 0.4145 & 0.1795 \\
         & Workday & \underline{0.1566} & \textbf{0.3600} & \textbf{0.4316} & \textbf{0.5081} & \underline{0.2519} & 0.1607 & \underline{0.3883} & \underline{0.4647} & \underline{0.5269} & 0.2628 & 0.1350 & 0.3418 & 0.4322 & 0.5074 & 0.2321 \\
         & Weekend & 0.0845 & 0.2153 & 0.2825 & \underline{0.3441} & 0.1499 & 0.0977 & 0.2295 & 0.2786 & 0.3282 & 0.1607 & 0.0757 & 0.2095 & 0.2750 & 0.3435 & 0.1405 \\
        \midrule
         & Local & \textbf{0.1629} & \underline{0.3206} & \underline{0.3824} & \underline{0.4393} & \underline{0.2386} & \textbf{0.2026} & \textbf{0.5500} & \textbf{0.6500} & \textbf{0.7184} & \textbf{0.3483} & 0.2076 & \textbf{0.4517} & \textbf{0.5243} & \underline{0.5910} & \underline{0.3175} \\
         & Tourist & 0.1059 & \textbf{0.2824} & \textbf{0.3570} & \textbf{0.4129} & \textbf{0.1885} & \textbf{0.1921} & \textbf{0.4868} & \textbf{0.5974} & \textbf{0.6316} & \textbf{0.3201} & 0.1402 & \textbf{0.3421} & \textbf{0.4228} & \textbf{0.4816} & \underline{0.2334} \\
         \bf{MSAHG} & Downtown & \underline{0.1871} & \underline{0.3680} & \underline{0.4349} & \underline{0.4882} & \underline{0.2723} & \textbf{0.2947} & \textbf{0.6974} & \textbf{0.8237} & \textbf{0.8605} & \textbf{0.4666} & 0.2132 & \underline{0.4858} & \underline{0.5689} & 0.6410 & \underline{0.3358} \\
         & Suburban & \textbf{0.1540} & \textbf{0.3151} & \textbf{0.3801} & \underline{0.4316} & \textbf{0.2310} & \textbf{0.1816} & \textbf{0.4447} & \textbf{0.5658} & \textbf{0.6000} & \textbf{0.3013} & \underline{0.1993} & \underline{0.4073} & \underline{0.4789} & \underline{0.5372} & \underline{0.2927} \\
         & Workday & \textbf{0.1882} & \underline{0.3585} & \underline{0.4301} & \underline{0.4945} & \textbf{0.2704} & \textbf{0.2316} & \textbf{0.5474} & \textbf{0.6526} & \textbf{0.7158} & \textbf{0.3736} & \underline{0.2171} & \textbf{0.4804} & \textbf{0.5620} & \underline{0.6216} & \underline{0.3351} \\
         & Weekend & 0.0960 & \underline{0.2250} & \underline{0.2857} & 0.3346 & 0.1580 & \textbf{0.1868} & \textbf{0.4395} & \textbf{0.5500} & \textbf{0.5763} & \textbf{0.3011} & \underline{0.1446} & 0.3218 & 0.3902 & 0.4534 & 0.2274 \\
        
        \bottomrule
    \end{tabular}}}
    \caption{Performance across multi-scenarios on the Gowalla, NYC, and TKY datasets is compared based on the Accuracy (Acc) and Mean Reciprocal Rank (MRR) metrics, with the best results highlighted in bold and the second-best results underlined.}
    \label{tab:main_comparison}
\end{table*}

\subsubsection{Baselines}
To demonstrate the performance of MSAHG, we implement five state-of-the-art (SOTA) methods as comparison baselines, covering both sequence-based and graph-based networks as follows:

$\bullet$  \textbf{HST-LSTM}~\cite{hst-lstm} integrates spatial-temporal factors into LSTM gates and models historical visits via a hierarchical encoder-decoder for location prediction.

$\bullet$  \textbf{DeepMove}~\cite{deepmove} considers long-term and short-term interests of users by combining recurrent neural networks with attention mechanisms.

$\bullet$  \textbf{GETNext}~\cite{getnext} introduces the global mobility patterns of all users into the Transformer architecture to improve model prediction effects.

$\bullet$  \textbf{STHGCN}~\cite{sthgcn} models trajectory relations via hypergraphs and integrates spatial-temporal information with a hypergraph transformer for next POI recommendation.

$\bullet$  \textbf{DCHL}~\cite{dchl} disentangles collaborative, transitional, and geographical views via hypergraphs, adapts fusion, and applies cross-view contrastive learning for next POI recommendation.

\subsubsection{Metrics}
Following STHGCN~\cite{sthgcn}, we use Acc@k (k=1, 5, 10, 20) and Mean Reciprocal Rank~(MRR) as evaluation metrics. Acc@k measures the proportion of true POIs in top-k predictions, assessing classification precision. MRR quantifies the average reciprocal rank of true POIs, evaluating ranking quality. Together, they comprehensively evaluate the model performance.

\subsubsection{Implementation details}
We implement \textit{MSAHG} using PyTorch 2.4.1 on a Linux server equipped with 503 GB RAM, a 160-core Intel\textsuperscript{\textregistered} Xeon\textsuperscript{\textregistered} Gold 6230 CPU (2.10 GHz), and A40 GPUs. The embedding dimensions for both POIs and users are set to 128, and the hypergraph convolutional network consists of three layers. Furthermore, we set 2.5km as distance threshold according to \cite{dchl}. The regularization weight $\lambda$ for the contrastive loss is configured to 0.1. 
To support stable parameter splitting (which resolves conflicting optimization across scenarios), we check for parameter conflicts after 20 epochs on the NYC and Gowalla datasets, and 30 epochs on TKY, given its larger data volume compared to the other two. Gradient similarities across scenarios are computed every 10 training batches to check the conflicts, and parameters are split into scenario-specific subsets when their similarities fall below -0.5.
The model is trained for 100 epochs using the Adam~\cite{kingma2017adammethodstochasticoptimization} optimizer with a learning rate of $1 \times 10^{-3}$, weight decay of $5 \times 10^{-4}$, and a batch size of 200, incorporating early stopping with a patience of 10 epochs. Baseline models are reproduced strictly following the hyperparameter settings specified in their original papers, and for any unspecified parameters, we adopt the same configurations as those of our proposed model.

\subsection{Experimental Results}
\subsubsection{Overall Performance.} 

The comprehensive evaluation across three real-world LBSN datasets demonstrates MSAHG's effectiveness for multi-scenario next POI recommendation. As shown in Table~\ref{tab:main_comparison}, MSAHG achieves state-of-the-art performance in 53.3\% of evaluation metrics (48/90 cases) and ranks second in 33.3\% (30/90), securing top-two positions in 86.7\% of all scenario-metric combinations.

\subsubsection{Analysis on Multi-Scenario Recommendation} 

This consistent performance across diverse scenarios is driven by two synergistic components. First, MSAHG captures scenario-specific features via sub-hypergraph construction, enabling precise modeling of contextual patterns that conventional methods overlook. Second, it mitigates inter-scenario conflicts through adaptive parameter splitting, which dynamically disentangles conflicting optimization signals and reduces learning interference.

These complementary mechanisms collectively ensure robust performance across all scenarios: the sub-hypergraphs enable discriminative modeling of scenario-specific mobility patterns, while the parameter splitting facilitates concurrent optimization of divergent features that would otherwise create optimization conflicts in joint training.

\subsubsection{Performance on Category Proportions}

\begin{figure}[t]
    \centering
    \includegraphics[width=1\columnwidth]{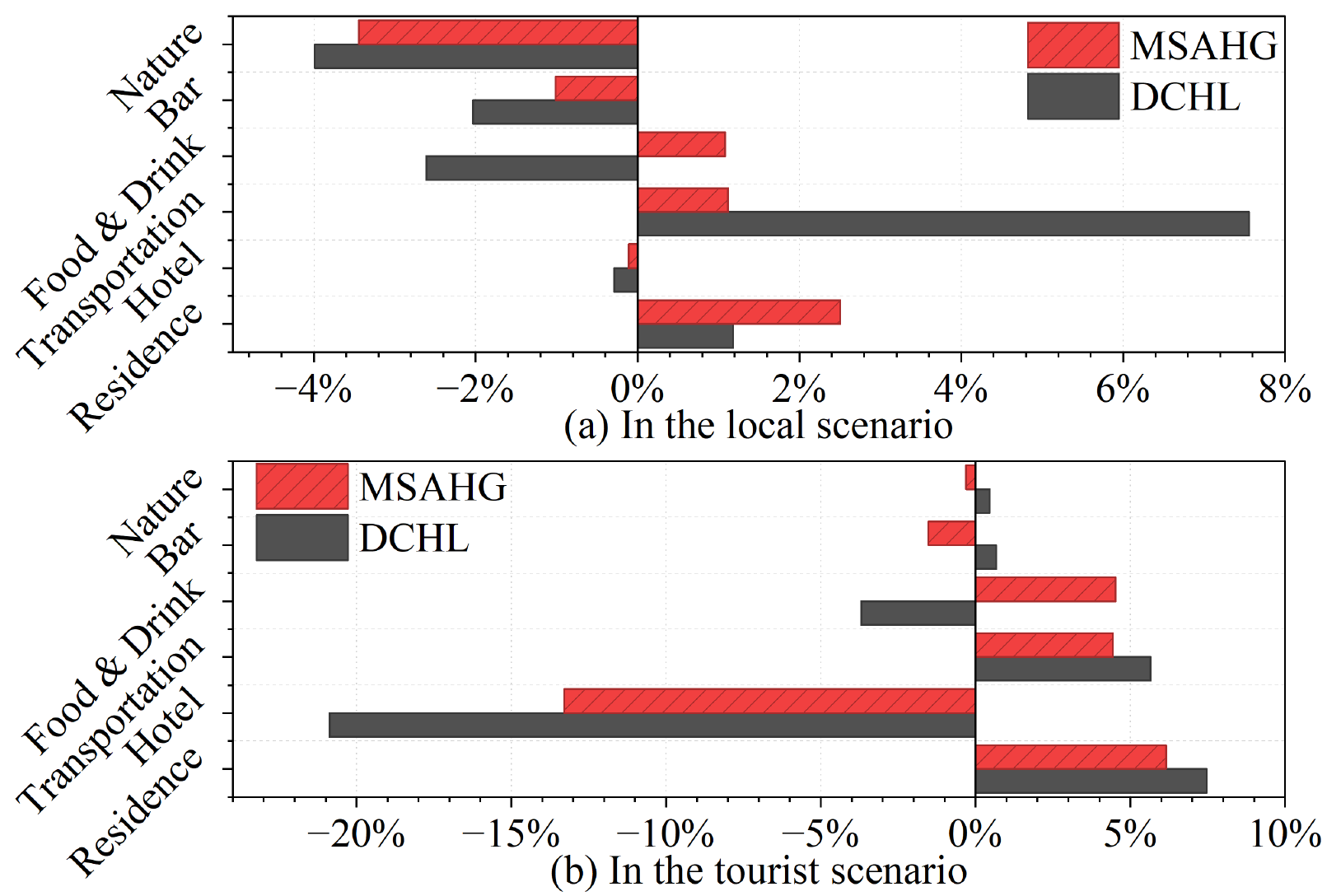}
    \caption{Comparison of category $\Delta$Percentage (difference between predicted and true category proportions) for DCHL and MSAHG in Local and Tourist scenarios on the NYC dataset. MSAHG better preserves the original distribution.}
    \label{fig:comparsion_categroy_proporation}
\end{figure}

Figure~\ref{fig:preliminaries_user_category} reveals distinct POI category distribution patterns between local and tourist scenarios, establishing characteristic differences that define scenario-specific behaviors. To validate MSAHG's ability to preserve these characteristics in predictions, we compare its generated POI category proportions against ground truth distributions and benchmark method DCHL in Figure~\ref{fig:comparsion_categroy_proporation}. Compared to DCHL, MSAHG aligns more closely with actual data distributions, particularly evident in Hotels and Transportation categories. While DCHL exhibits significant deviation in tourist scenarios - indicating failure to capture distinctive scenario patterns. This contrast confirms MSAHG's capacity to model scenario-specific feature patterns.\looseness=-1

\subsubsection{Performance on Distance Distribution}

\begin{figure}[t]
    \centering
    \includegraphics[width=1\columnwidth]{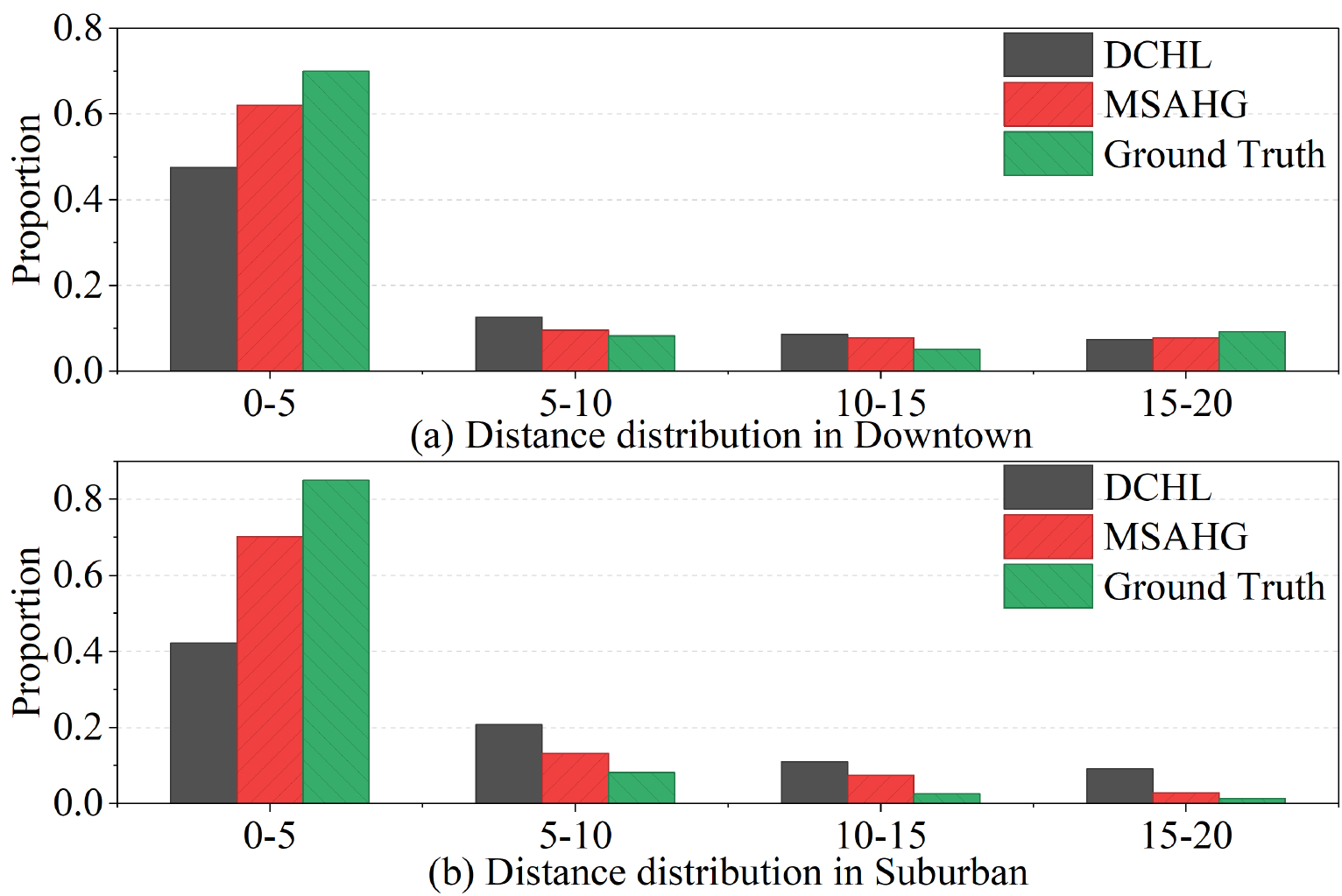}
    \caption{Comparison of POI distance distribution performance between DCHL and MSAHG in Downtown and Suburban scenarios on TKY dataset.}
    \label{fig:distance_distribution}
\end{figure}

Beyond category distribution patterns, we observe significant differences in inter-POI distances across scenarios. Figure~\ref{fig:distance_distribution} compares distance distributions of predicted check-ins by MSAHG and DCHL against ground truth in downtown and suburban scenarios on the TKY dataset. MSAHG consistently matches actual distance distributions more closely than DCHL in both scenarios. This advantage is especially evident in suburban contexts: while real-world data shows 85\% of check-ins occur within 0-5km, MSAHG accurately captures this pattern with 70\% of predictions in this range. In contrast, DCHL predicts only 42\% - nearly matching its downtown distribution rather than adapting to suburban patterns. This indicates DCHL fails to recognize distinctive suburban spatial characteristics. Simultaneously, MSAHG accurately preserves downtown-specific long-distance patterns (15-20 km), which fundamentally differ from suburban distributions. This confirms MSAHG's capability to simultaneously model divergent spatial patterns across scenarios, capturing both suburban and downtown mobility.

\subsection{Ablation Study}
To evaluate component effectiveness in MSAHG, we conduct ablation studies on the NYC dataset comparing the full model against two variants: (1) without parameter splitting and (2) without scenario-specific sub-hypergraphs. 
\begin{table}[htbp]
\scriptsize 
\centering
\renewcommand{\arraystretch}{0.7} 
\resizebox{\columnwidth}{!}{
    \begin{tabular}{@{}lccccc@{}}
    \toprule
    \textbf{Model} & \textbf{ACC@1} & \textbf{ACC@5} & \textbf{ACC@10} & \textbf{ACC@20} & \textbf{MRR} \\
    \midrule
    \makecell[l]{$w/o$ \\ split parameter} & 0.2026 & 0.5270 & 0.6416 & 0.6826 & 0.3392 \\
    \midrule[0.4pt]
    \makecell[l]{$w/o$ \\ sub-hypergraph} & 0.2164 & 0.5417 & 0.6514 & 0.6965 & 0.3592 \\
    \midrule[0.4pt]
    Full Model & \textbf{0.2268} & \textbf{0.5605} & \textbf{0.6743} & \textbf{0.7191} & \textbf{0.3721} \\
    \bottomrule
    \end{tabular}
}
\caption{Ablation study results over NYC dataset.}
\label{tab:ablation_nyc_scenarios}
\end{table}

As shown in Table \ref{tab:ablation_nyc_scenarios}, both components contribute significantly to performance. Specifically, the parameter-splitting mechanism exerts a notable impact by enabling the model to assign independent optimization directions to conflicting scenarios, thereby facilitating the concurrent learning of mobility patterns across scenarios. Additionally, incorporating scenario-specific sub-hypergraphs further enhances performance, as this allows the model to capture trajectory features under different scenarios. Collectively, the full model outperforms both variants across all evaluation metrics, confirming that these components complement each other in modeling complex multi-scenario behaviors.





\subsection{Computational Efficiency}


\begin{table}[tbp]
\scriptsize 
\centering
\renewcommand{\arraystretch}{0.7} 
\setlength{\tabcolsep}{2pt} 
\resizebox{\linewidth}{!}{
    \begin{tabular}{@{}l c c c c@{}}
    \toprule
    \textbf{Model} & \textbf{Train Time (s)} & \textbf{Train Mem (MB)} & \textbf{Test Time (s)} & \textbf{Test Mem (MB)} \\
    \midrule
    MSAHG   & 10.69   & 1,292.36 & 12.47 & 224.61 \\
    \midrule[0.4pt]
    \makecell[l]{MSAHG \\ (w/o split\&subgraph)} & 10.26 & 697.66 & 11.49 & 225.35 \\
    \midrule[0.4pt]
    STHGCN  & 73.29   & 672.24   & 8.68  & 639.85 \\
    \midrule[0.4pt]
    GETNext & 408.92  & 3,714.18 & 6.30  & 1,313.91 \\
    \bottomrule
    \end{tabular}
}
\caption{Efficiency comparison on NYC dataset. Metrics are per epoch (single GPU).}
\label{tab:efficiency_comparison}
\end{table}
To evaluate the computational efficiency of sub-hypergraph and adaptive parameter splitting modules, we benchmark MSAHG against an ablation variant and two baseline models, as shown in Table \ref{tab:efficiency_comparison}. The results demonstrate that MSAHG is highly time-efficient. While the high training memory consumption stems from gradient-similarity buffers rather than an increase in model parameters, a fact supported by its lower memory footprint during testing.


\subsection{Hyper-parameter Study}

We conduct a comprehensive sensitivity analysis on four key hyperparameters: the balancing coefficient~\(\lambda\), stacking hypergraph convolutional layers, temperature~\(\tau\), and divide epoch across scenarios. The experiment results reveal that MSAHG maintains robust performance across diverse hyperparameter settings. Detailed results are provided in the Appendix.

\section{Conclusion}


We propose MSAHG for multi-scenario POI recommendation, which captures distinct trajectory features across scenarios via a scenario-splitting paradigm, with its efficacy validated experimentally.
Our future work will explore inter-scenario relationships to enhance shared feature learning, thereby further boosting multi-scenario performance.

\section{Acknowledgments}
This research was funded by the Science and Technology Development Program of Jilin Province (20250203122SF) and China Scholarship Council (202308440220).
\bibliography{aaai2026}

\clearpage
\section{Appendix}
\subsection{Dataset}
\label{dataset}
The NYC and TKY datasets are both derived from Foursquare, with NYC encompassing POIs within a 40km radius of New York City’s center and TKY including those within a 40km radius of Tokyo’s center. In contrast, the Gowalla dataset consists of check-in records spanning from February 2009 to October 2010, with coverage focused on the regions of California and Nevada.  

\subsubsection{Multiple Scenario Categorizing}  
To enable multi-scenario analysis, we categorized the data across three dimensions:  

\begin{itemize}
    \item \textbf{User type}: Users are categorized as either locals or tourists. Specifically, tourists are defined as users whose accommodation-related check-ins exceed 5\% of their total check-ins.
    
    \item \textbf{Spatial region}: POIs are classified into downtown (within 10 km of the city center) or suburban (beyond 10 km) areas. For Gowalla, city centers are identified in cities with populations over 100,000, as sparsely populated areas tend to exhibit suburban characteristics.
    
    \item \textbf{Temporal period}: Trajectories are labeled as occurring on a workday or weekend based on the timestamp of the penultimate check-in (the one preceding the target check-in), accounting for the distinct behavioral patterns observed across these periods.
\end{itemize}

We chronologically split each user's check-in records, with 80\% for training and 20\% for testing.  For scenario-specific grouping, trajectories are categorized based on the last observed location and timestamp, excluding target check-in information. This ensures alignment with real-world prediction scenarios where future context is unavailable during inference.

Table~\ref{tab:data-statistics} presents detailed statistics of the datasets, including the number of users, POIs, check-ins, and trajectories, along with their distributions across the multi-scenario categories.

\begin{table}[!ht]
    \centering
    \resizebox{\linewidth}{!}{
    \begin{tabular}{@{}lccc@{}}
        \toprule
        \textbf{Metric}              & \textbf{Gowalla} & \textbf{NYC} & \textbf{TKY} \\
        \midrule
        
        \multicolumn{4}{@{}l@{}}{\textit{Basic Statistics}} \\[3pt] 
        Users                        & 3,996            & 1,743        & 1,383        \\
        POIs                         & 9,823            & 7,289        & 17,023       \\
        Check-ins                    & 238,488          & 57,327       & 507,260      \\
        Trajectories                 & 24,499           & 6,102        & 41,374       \\ 
        
        \midrule[0.5pt] 
        
        \multicolumn{4}{@{}l@{}}{\textit{Multi-scenario Categories}} \\[3pt] 
        Local Users                  & 3,219            & 1,146        & 1,259        \\
        Tourists              & 777              & 597          & 124          \\
        Downtown POIs                & 7,229            & 5,291        & 11,742       \\
        Suburban POIs                & 2,603            & 1,998        & 5,281        \\
        Workday Trajectories         & 15,706           & 4,174        & 28,792       \\
        Weekend Trajectories         & 8,793            & 1,928        & 12,582       \\
        \bottomrule
    \end{tabular}}
    \caption{Dataset and Multi-scenario Statistics}
    \label{tab:data-statistics}
\end{table}

\subsection{Hyper-parameter Study}

\begin{figure}[t]
    \centering
    \includegraphics[width=1\columnwidth]{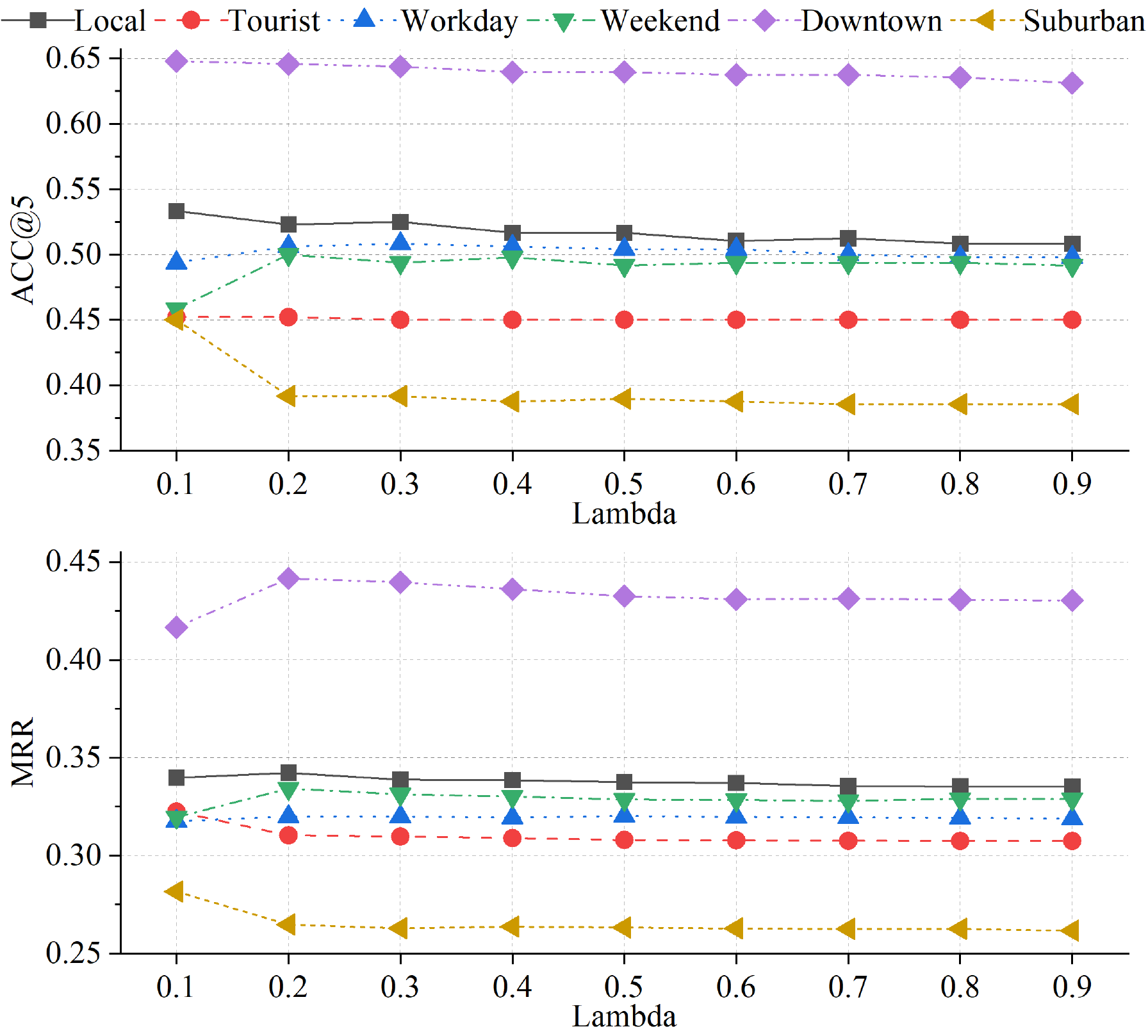}
    \caption{Hyper-parameter sensitivity study on Lambda.}
    \label{fig:lambda_hyper}
\end{figure}

\begin{figure}[t]
    \centering
    \includegraphics[width=1\columnwidth]{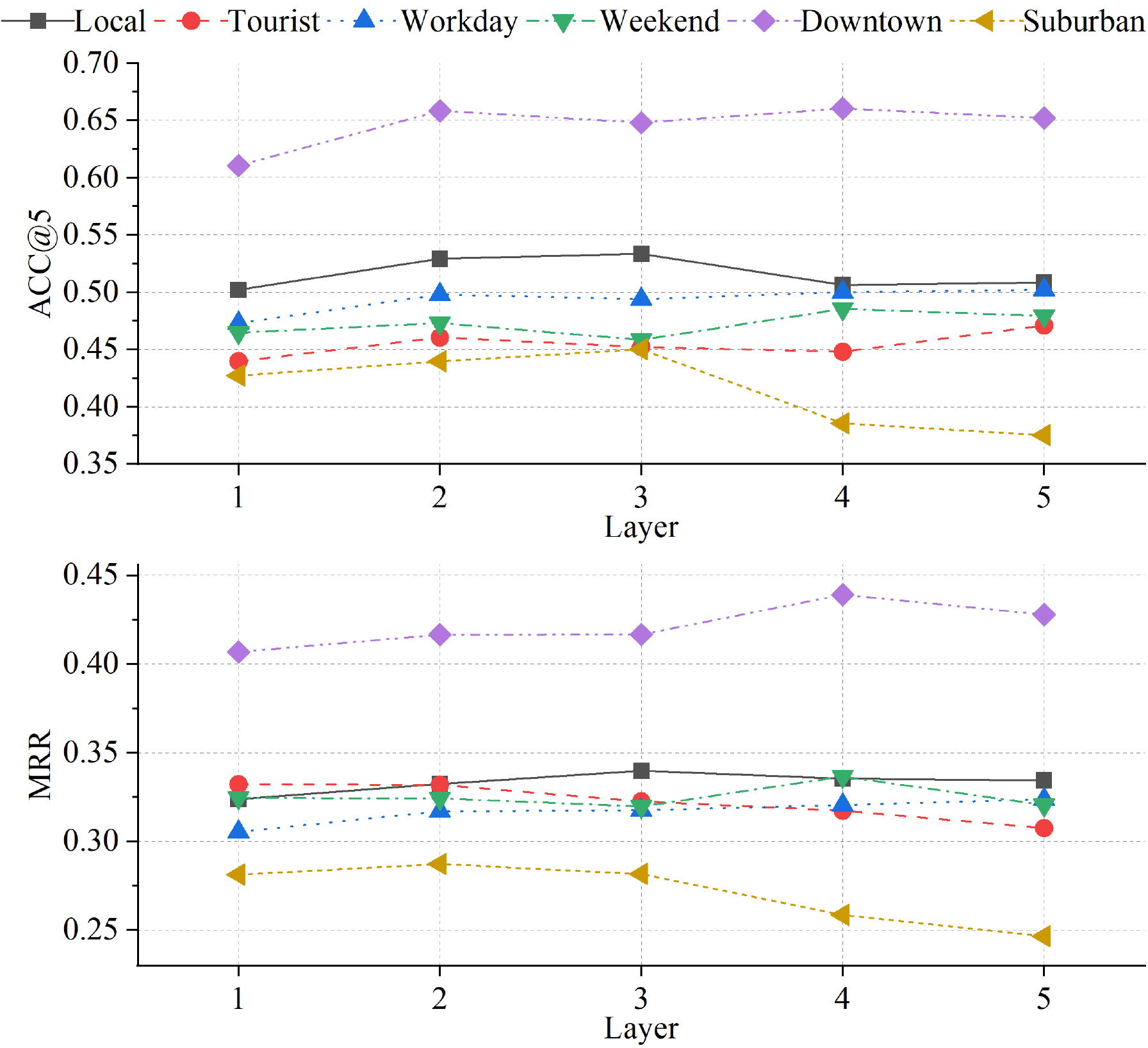}
    \caption{Hyper-parameter sensitivity study on Layer.}
    \label{fig:layer_hyper}
\end{figure}

We conduct a comprehensive sensitivity analysis of MSAHG on the NYC dataset, examining the impact of four key hyperparameters—\(\lambda\), stacking hypergraph convolutional layers, temperature~\(\tau\), and divide epoch—using ACC@5 (measuring the proportion of true Points of Interest (POIs) in the top-5 predictions) and MRR (quantifying the average reciprocal rank) as evaluation metrics.

\subsubsection{Hyper-parameter Study on Lambda} 
For the hyperparameter~\(\lambda\), which balances constructive and recommendation losses, was evaluated across $\lambda \in \{0.1, 0.2, \dots, 0.9\}$. As depicted in Figure~\ref{fig:lambda_hyper}, both ACC@5 and MRR metrics demonstrate stability throughout the tested range. While suburban regions achieve optimal performance at $\lambda = 0.1$ before plateauing, minimal fluctuations occur in other scenarios. This consistent behavior across diverse contexts confirms the model's robustness to variations in loss function weighting.

\subsubsection{Hyper-parameter Study on Layer} 
The impact of hypergraph convolutional depth was examined through layer configurations $\{1, 2, 3, 4, 5\}$. Figure~\ref{fig:layer_hyper} reveals a characteristic performance trajectory: ACC@5 and MRR progressively improve from 1 to 3 layers, peak at the 3-layer configuration, then exhibit degradation beyond this depth. This pattern suggests a balance between signal propagation and noise introduction - shallower architectures may insufficiently capture high-order relationships, while excessive layers potentially propagate irrelevant signals that dilute feature quality. The moderate performance variance across layers further underscores the architecture's resilience to depth adjustments.
\begin{figure}[t]
    \centering
    \includegraphics[width=1\columnwidth]{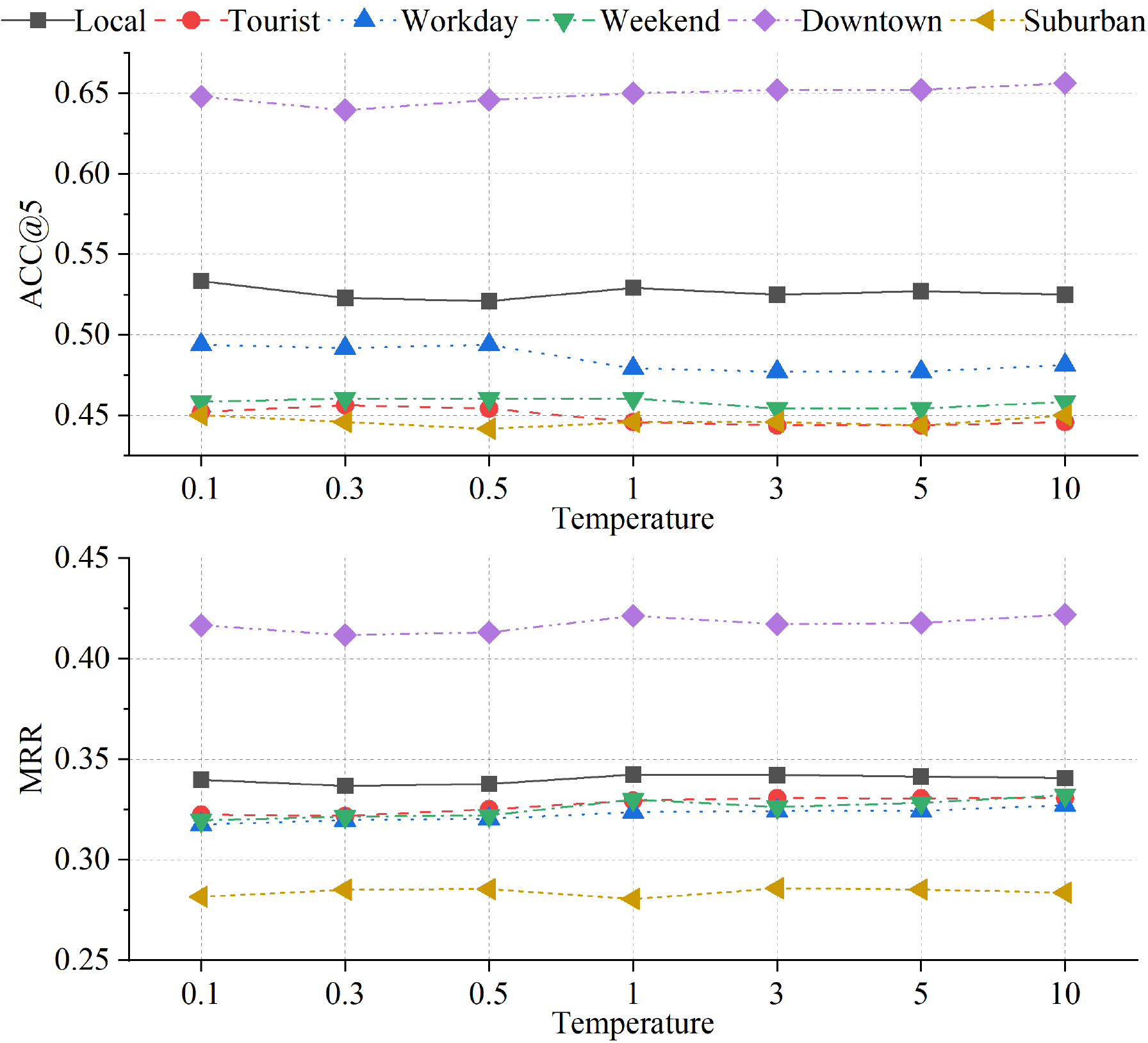}
    \caption{Hyper-parameter sensitivity study on Temperature.}
    \label{fig:tempture_hyper}
\end{figure}

\begin{figure}[t]
    \centering
    \includegraphics[width=1\columnwidth]{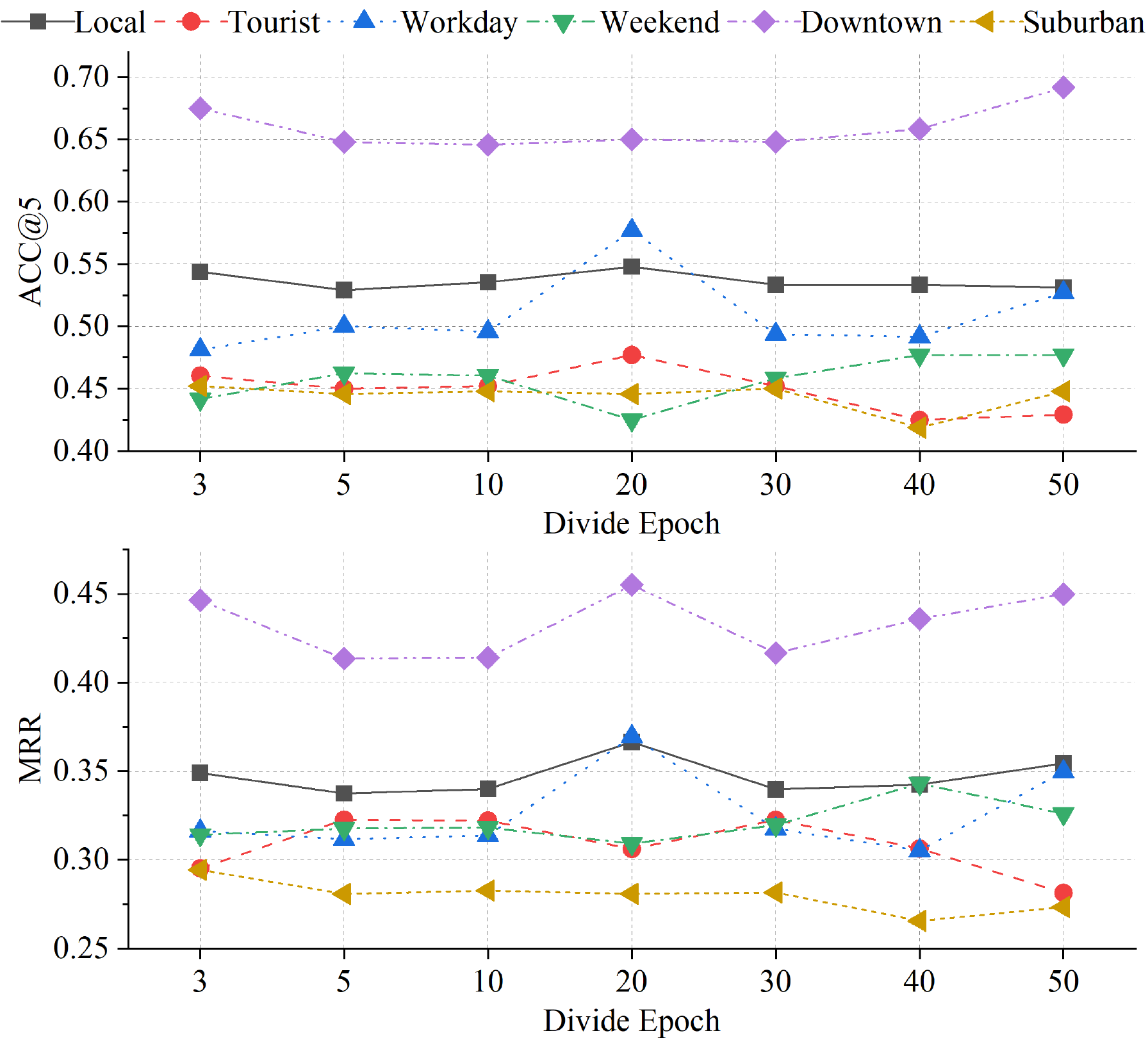}
    \caption{Hyper-parameter sensitivity study on divide epoch.}
    \label{fig:divide_epoch_hyper}
\end{figure}

\subsubsection{Hyper-parameter Study on Temperature}
For the temperature~\(\tau\), which controls the strength of gradients in contrastive learning, we test values~\(\tau \in \{0.1, 0.3, 0.5, 1, 3, 5, 10\}\). Figure~\ref{fig:temperature_hyper} shows that the MSAHG model remains stable as ~\(\tau\) varies. These results confirm the model’s robustness to contrastive learning temperature fluctuations.

\subsubsection{Hyper-parameter Study on Divide Epoch}
The divide epoch denotes the number of epochs trained prior to applying the divide conflict parameter, with its values adjusted according to dataset sizes (e.g., 20 epochs for NYC/Gowalla and 30 for TKY). For the NYC dataset, we test divide epoch values in \(\{3, 5, 10, 20, 30, 40, 50\}\), as shown in Figure~\ref{fig:divide_epoch_hyper}. The results demonstrate that our framework maintains remarkable performance stability across a wide range of values. Notably, the model performs robustly even at 3 epochs. This suggests that early-stage parameter splitting effectively leverages nascent feature representations before full convergence, enabling the model to cultivate scenario-specific knowledge more rapidly. As the training progresses to intermediate epochs, performance plateaus at a high level. At this stage, the model has achieved sufficient fitting, allowing the mechanism to precisely identify and decouple parameters with genuine cross-scenario conflicts.Ultimately, the consistent performance across different settings underscores the robustness of our adaptive splitting framework and its efficacy in accommodating diverse data distributions.

Overall, these experiments confirm that MSAHG maintains robust performance across diverse hyper-parameter settings, with only minor fluctuations in specific scenarios. This stability underscores the model’s ability to generalize under varying architectural and training configurations.

\end{document}